# Noncontact measurement method of linear and angular displacement based on dual-beam feedback interferometric system

Xin Xu, Zongren Dai, and Yidong Tan

*Abstract*— **This study describes a unique optical approach for the noncontact measurement of linear and angular displacement. Compared to previous methods, the sensor system here based on the dual-beam phase-modulated feedback interferometry provides higher sensitivity for non-cooperative targets and a wider range concerning the angle measurement. The amount of linear and angular displacement is calculated by tracing the phase changes of the differential beams. Performance of the proposed method is evaluated via testing a prototype system. The prototype has a 35 nm and 0.15" stability over 1 hour, with a resolution of 1 nm and 0.02" correspondingly, according to the experimental data. The linearity is $5.58\times10^{-6}$ in the range of 100 mm and $1.34\times10^{-4}$ in the range of 360°, indicating that the proposed method may possess considerable potential for high-precision metrological applications.**

*Index Terms*—**Linear and angular displacement, noncontact measurement, feedback interferometry.**

## I. INTRODUCTION

IN industrial applications and frontier scientific research, such as navigation, robotics, machine positioning and mask aligner [1-5], linear and angular displacement measurements are critical. High precision with a large sensing range is the mutual requirement of the incremental linear or angular encoders for these applications. For example, the ongoing space gravitational detection mission seeks to detect a displacement sensitivity of 1 pm/Hz$^{1/2}$ and 1 nrad/Hz$^{1/2}$ in the low-frequency band at the kilometer-level distance [6-8]. Other elements like compactness, stability and repeatability are also important in various applications using linear or angular displacement sensors [9]. Over the past several decades, different types of linear/angular displacement sensors with specific benefits and limitations to each other have been reported, including mechanical, electromagnetical and optical methods. Mechanical sensors to measure position or angle are commercially available, but the micrometer-level accuracy and resolution limit its further applications. An electromagnetical incremental encoder is a communication device that translates linear or rotary motion into a digital signal with respect to capacitance/inductance/resistance/potential change [10-16]. Owing to the features such as high accuracy, small size and low power consumption, the electromagnetical encoders have a well-established performance history, which provide sub-micrometer accuracy in the range of several millimeters or arcsec accuracy in the 360° range [13-16]. Boby, et al have well summarized the advantages and disadvantages of capacitive, inductive and magnetic displacement sensors [12]. Despite their variety and good performance, most electromagnetical sensors need to connect with the moving or rotating targets during the measurement process. This feature may disrupt the motion state of the system and cause sensing mistakes, thereby limiting the application possibilities. Furthermore, sensor systems able to simultaneously measure displacement and rotation are still rare [13].

Because of its flexibility, nondestructive detection, high precision and resolution, optical sensors evolve as quite attractive alternatives to electromagnetical sensors since the invention of lasers [17-26]. Heterodyne interferometry is one of the most precise methods to measure linear and angular displacement through tracing the phase variation [6,17-18]. A dual-heterodyne laser interferometer with a sensitivity noise level of 1 pm/Hz$^{1/2}$ and 0.4 nrad/Hz$^{1/2}$ at 1 Hz was used to detect linear and angular displacement simultaneously in 2019 [19]. Another representative method based on optical grating can act as the standard of length and rotation variation [20-23]. Nevertheless, no matter the cooperative mirrors in the interferometric sensors or the installation of the grating to the moving target will bring inconvenience and potential errors. Some other optical methods can also measure the incremental displacement or rotation, including Faraday effect [24], photodetectors [25-26], dual-comb interferometry [27], fiber interferometry [28-29]. The above methods have their own advantages such as high precision, high resolution and multi-freedom measurement. However, building a sensor system for noncontact motion measurement without cooperating mirrors, particularly one with a higher angular detection range, can be difficult and prohibitive.

Recently, laser feedback interferometry has attracted much attention due to its self-collimation, compactness, noncontact sensing and ultra-high sensitivity under frequency-shift

Manuscript received Month xx, 2xxx; revised Month xx, xxxx; accepted Month x, xxxx. This work was supported by the National Key Research and Development Program of China (No. 2020YFC2200204), National Outstanding Youth Science Fund Project of National Natural Science Foundation of China (No. 51722506) and Equipment Development Department Project. (*Corresponding author: Yidong Tan*).

The authors are with the Department of Precision Instruments, Tsinghua University, Beijing 100084, China (email: xx19@mails.tsinghua.edu.cn; dzr18@mails.tsinghua.edu.cn; tanyd@tsinghua.edu.cn)

modulation, such as vibration/displacement/velocity measurement, biomedical imaging and thermal sensing [30-38]. However, few studies have applied feedback interferometry to measure both linear and angular displacement [35,38]. Therefore, based on the unique features of laser feedback interferometry, the ambition of this paper is to present an optical sensor system that is capable of noncontact measurement for linear and angular displacement without mirrors. While it enables high sensitivity for the measurement of different targets, it also offers the advantages of millimeter-level displacement and full-circle rotation sensing range. The paper is organized into four main sections. First, the optical design and working principle are introduced in Section II. Signal processing and linear/angular displacement calculation are also presented in this part. Then, a prototype sensor system is built and tested, with which the performance results of stability, resolution, linearity and repeatability are shown in Section III. In section IV, error analysis and extended applications of the proposed system are discussed. The conclusion is summarized in Section V.

## II. DESIGN AND WORKING PRINCIPLE

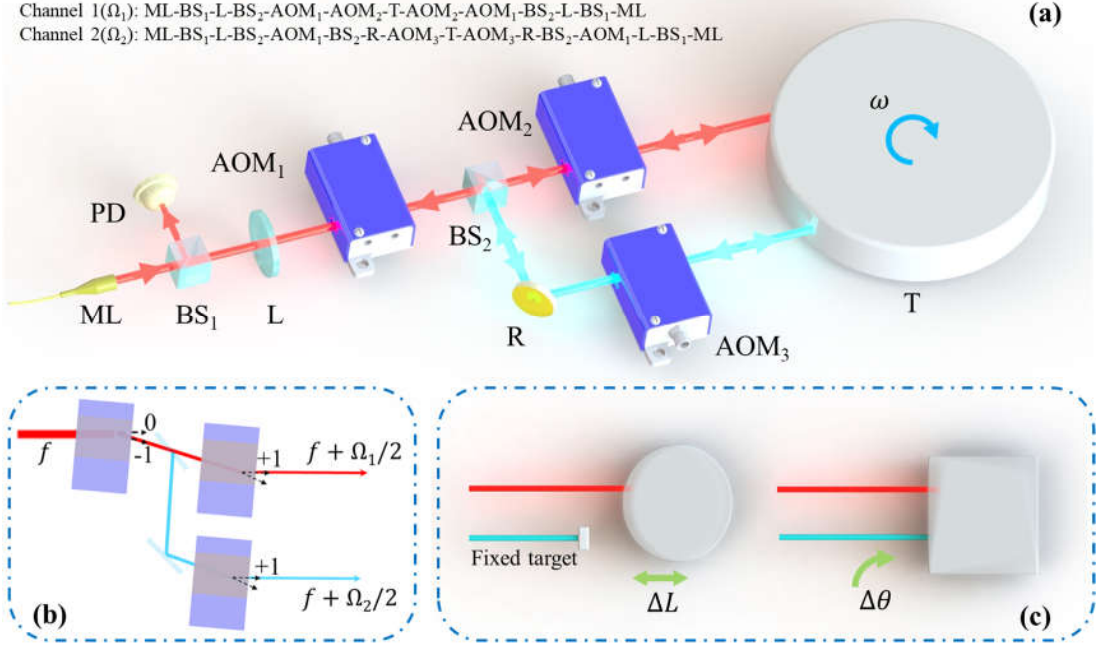

Fig. 1. Diagram of the optical sensor for linear and angular measurement. (a) Optical setup. (b) Heterodyne modulation (c) The beam illustration when the target is moving or rotating.
ML: microchip laser; PD: photodetector; BS: beam splitter; L: lens; AOM: acoustic-optic modulator; R: reflector; T: target;

### A. Optical design

A structural diagram of the proposed sensor for noncontact measurement of the linear and angular displacement is shown in Fig. 1. A solid-state microchip laser is employed as the sensor source with the fundamental transverse mode and single longitudinal mode output [39]. The output beams is measured 1064.24 nm using a wavelength meter (HighFinesse, WS 867). As Fig. 1(a) illustrates, the output beams are divided into two parts by a beam splitter (BS$_1$). The transmissive part is modulated by acoustic-optic modulators, the center frequency of which are 70MHz, 70.8MHz and 71.5MHz, respectively. The detailed frequency-shift modulation is shown in Fig. 1(b). $f$ is the frequency of the output beams. To achieve ultra-high gain, the system utilizes a differential heterodyne structure, in which the round-shifted modulation is 1.6MHz ($\Omega_1$) and 3MHz ($\Omega_2$) for the two sensing beams [30]. The specific light paths of two sensing beams are also listed in Fig. 1(a). A fixed target is placed as a reference when the system is employed for the measurement of linear displacement $\Delta L$. It can also be seen that the system is suitable for noncontact measurement of linear displacement $\Delta L$ and angular displacement $\Delta \theta$, no matter the target is cylindrical or cuboid (see Section II Part B for a detailed description). The reflective part of the output beams containing the modulated signals is received by an infrared photodetector. The intensity modulation under frequency shift can be described as [30, 34]:

$$\frac{\Delta I(\Omega_1)}{I} = \kappa_1 G(\Omega_1) \cos(2\pi\Omega_1 t - \phi_{01} + \Delta\phi_1) \quad (1)$$

$$\frac{\Delta I(\Omega_2)}{I} = \kappa_2 G(\Omega_2) \cos(2\pi\Omega_2 t - \phi_{02} + \Delta\phi_2) \quad (2)$$

where κ is the effective reflection coefficient of the external target, $\phi_{01}$ and $\phi_{02}$ are the initial phase of $\Omega_1$ and $\Omega_2$ channel, respectively. G is the gain function. $\Delta\phi_1$ and $\Delta\phi_2$ represent the external phase of the two different channels. When the target is moving or rotating, the phase items will vary accordingly, thus providing a method to trace the linear and angular displacement.

### B. Sensing principles

The sensing principle for linear and angular displacement is illustrated in Fig. 2. When the target is moving at a velocity $v$, as Fig. 2(a) shows, it will cause a doppler effect on the incident beams. The Doppler frequency shift can be expressed as:

$$\Delta f_1 = 2f \frac{v\cos\theta}{c} \quad (3)$$

where $\theta$ is the angle between the beam propagation direction and the moving direction. $c$ is the speed of light, which is far greater than $v$ under normal circumstances.

As a fact, the integral of angular frequency over time is the phase change, thus the phase variation can be derived as:

$$\Delta\phi_1 = 2\pi \int \Delta f_1 \, dt = \frac{4\pi f}{c} \int v\cos\theta \, dt = \frac{4\pi \Delta L_1}{\lambda} \quad (4)$$

where $\Delta L_1$ represents the displacement along the $\Omega_1$ sensing beam and $\lambda$ is the wavelength of the output laser. The same derivation holds for the $\Omega_2$ sensing beam, as it is incident on a fixed target of quasi-common path, thus offering a reference compensation. Therefore, the linear displacement $\Delta L$ is linked to the phase variation of two parallel beams.

$$\Delta L = \frac{\lambda}{4\pi}(\Delta\phi_1 - \Delta\phi_2) \quad (5)$$

As to the measurement of angular displacement $\Delta\theta$, the derivation process is similar with the linear part, with a principle illustration shown in Fig. 2(b). $D$ is the distance between two parallel beams, which is 68.60 mm measured by the combination of a position-sensitive detector (50 μm precision and 10 μm resolution) and a high-precision linear stage (PI, Inc. M511, ±0.2 μm repeatability) in the experiments. When the target is rotating, the phase of two sensing beams can be expressed as:

$$\Delta\phi_1 = \frac{4\pi f}{c} \int v\cos\theta_1 \, dt = \frac{4\pi}{\lambda} \int \omega R\cos\theta_1 \, dt \quad (6)$$

$$\Delta\phi_2 = \frac{4\pi f}{c} \int v\cos\theta_2 \, dt = \frac{4\pi}{\lambda} \int \omega R\cos\theta_2 \, dt \quad (7)$$

where $\omega$ is the angular velocity and $R$ is the rotary radius. $\theta_1$ and $\theta_2$ are the angle between two beams and the line velocity, respectively. Based on the geometric relationship ($D = R\cos\theta_2 - R\cos\theta_1$) as shown in Fig. 2b, we subtract (6) from (7), and obtain the incremental angular displacement $\Delta\theta$ as:

$$\Delta\theta = \int \omega \, dt = \frac{\lambda}{4\pi} \frac{(\Delta\phi_2 - \Delta\phi_1)}{R\cos\theta_2 - R\cos\theta_1}$$

$$= \frac{\lambda}{4\pi D}(\Delta\phi_2 - \Delta\phi_1) \quad (8)$$

When the system suffers the environmental perturbation such as thermal disturbance or external vibration, the phase of two beams vary in the same direction. It means that the proposed angular measurement method also possesses a compensation for the environmental perturbation in a subtraction processing. Owing to the ultra-high gain of frequency-shift feedback interferometry, the sensor system can respond stably and continuously to the sensing beams scattered from the moving or rotating targets. Thus, a sensor employing feedback interferometry and differential dual-beam structure is successfully proposed. The linear and angular displacement can be traced through the phase variations of the system.

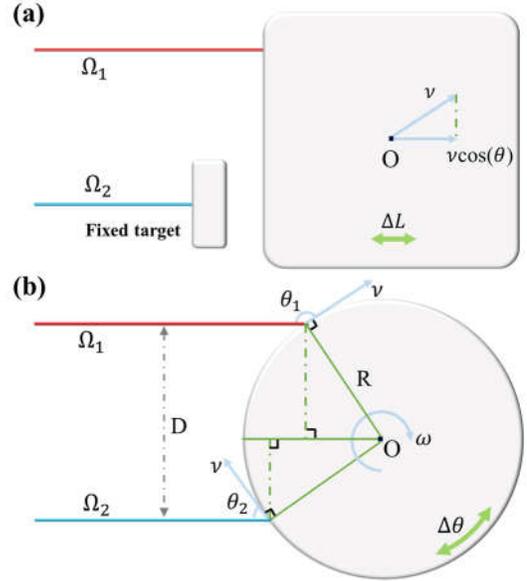

Fig. 2. Sensing principle illustration. (a) Measurement of the linear displacement. (b) Measurement of the angular displacement.

### C. Signal processing

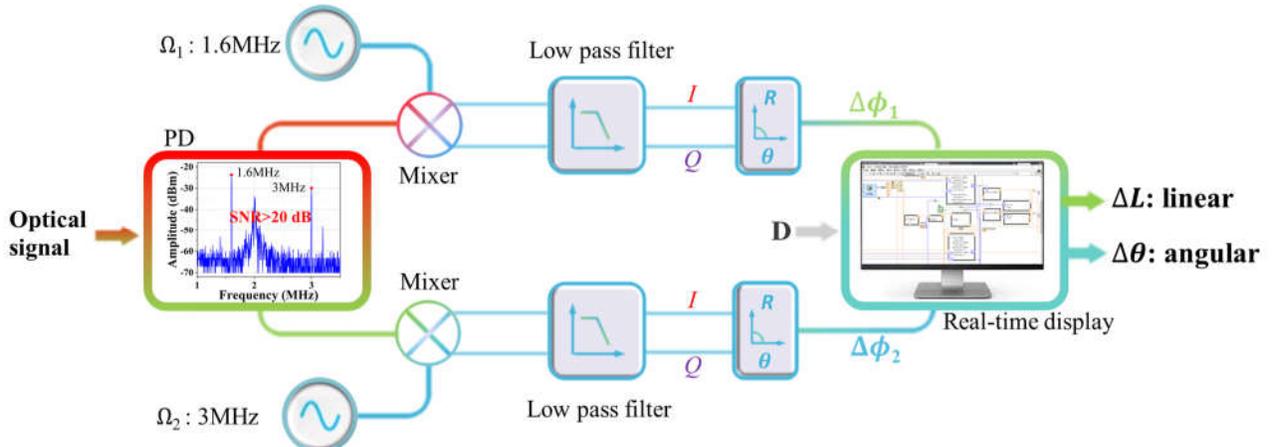

Fig. 3. Flow chart showing the demodulation process of the linear and angular displacement. $D$ is the distance between two parallel beams. $I$ is the in-phase component and $Q$ is the quadrature component. $R$ and $\theta$ are the amplitude and phase of the signal respectively. $\Delta\phi_1$ and $\Delta\phi_2$ represent the external phase of the two different channels.

Based on the optical design and sensing principle illustration, it is found that how to extract exactly the phase variations from the modulated intensity signals is the key step for the successful construction of such a linear/angular displacement sensor. In former studies, fringes counting is used to acquire displacement or angle information [35]. The moving/rotating velocity can also be fast obtained through spectrum analysis [36]. In this paper, In-phase Quadrature (IQ) demodulation method [40] is utilized to recover the phase changes of the modulated signals accurately, as Fig. 3 shows. The modulated beams, which are described by (1) and (2), are observed by an oscilloscope, showing that the SNR (signal-to-noise ratio) of both $\Omega_1$ and $\Omega_2$ channel exceed more than 20dB. The original signals are mixed by the references and their quadrature values, to be specific, 1.6 MHz and 3 MHz RF signals from the AOM drivers respectively. The high-frequency components are filtered and I/Q items are operated by the arctangent function to get the phases of two channels. Based on the lock-in amplifier (Zurich Instrument, HF2LI) and Labview operation, the linear and angular displacement can be calculated and displayed in real time. This method owns the advantages of precise traceability of the phase variations from unexpected noises, thus promising a high accuracy for the linear and angular displacement measurement.

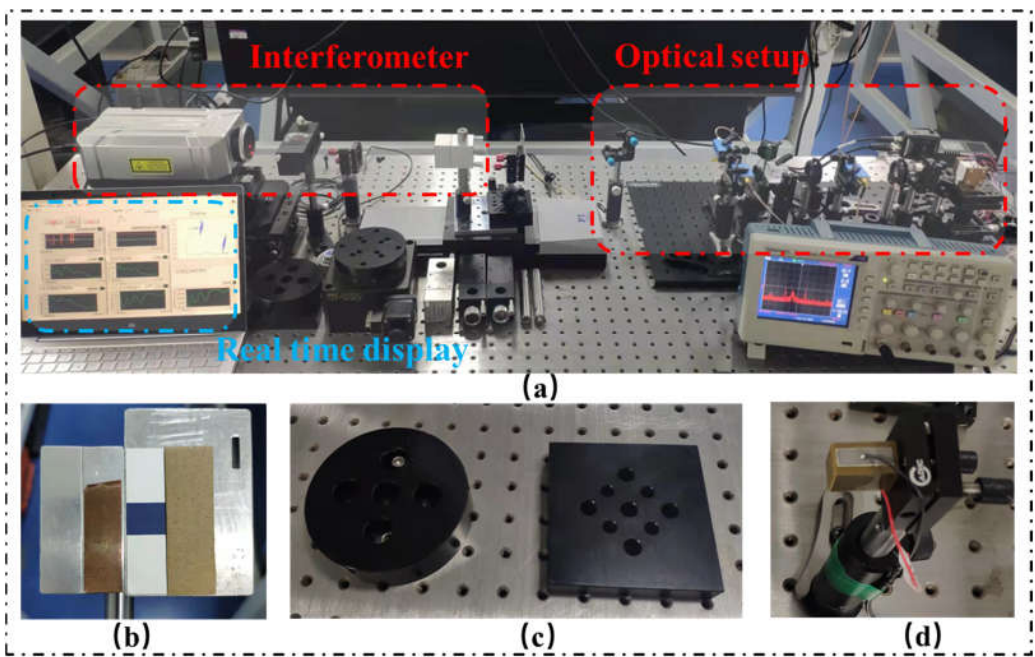

Fig. 4. The system for linear and angular displacement measurement. (a) Compared with the interferometer LeiCe 3000H. (b) The test targets for linear measurement made of aluminum foil/Copper foil/rough paper. (c) The test targets for angular measurement with a smooth aluminum surface. (d) The PZT actuator (XMT150).

### III. Performance test and results

#### A. Prototype sensor system

To verify the feasibility and effectiveness of the proposed method, a prototype has been built on a 300*600*13 mm breadboard, as shown in Fig. 4(a) with a commercial interferometer for comparison. Fig. 4(b) shows the test target for the measurement of linear displacement and Fig. 4(c) are the test targets for the measurement of angular displacement. The operating environment is in the normal laboratory without artificial temperature control and an acrylic cover has been used to decrease the airflow effect. The initial experiments prove that all the targets have a frequency-modulated response under 0.5 m working distance, whose SNR reach more than 20dB. The performance of the prototype sensor system is then tested via measuring its stability, resolution, linearity and repeatability.

#### B. Performance results

*1. Stability*

Stability describes the steady state of a system under no motion. To test the stability of the prototype, the rough paper is used as the test target for the linear displacement, and the reference is a common mount shown in Fig.4(a). The long-term stability result is presented in Fig.5(a), indicating that the drift within 1 hour is lower than 35nm. Note that $S_1/S_2$ are the respective linear displacement of two sensing beams. Then the stability test of angular displacement has been carried out by fixing the aluminum cylinder on the rotary stage, and the result is shown in Fig. 5(b). The built prototype has a stability of 0.15" over 1 hour.

The Noise Spectrum Density (NSD) results based on 1-hour linear/angular displacement drift are also given in Fig. 5(c) to find the sensitivity limit [17]. The spectrum density of linear and angular displacement can reach up to 0.05nm/Hz$^{1/2}$@1Hz and 0.9 nrad/Hz$^{1/2}$@1Hz, respectively.

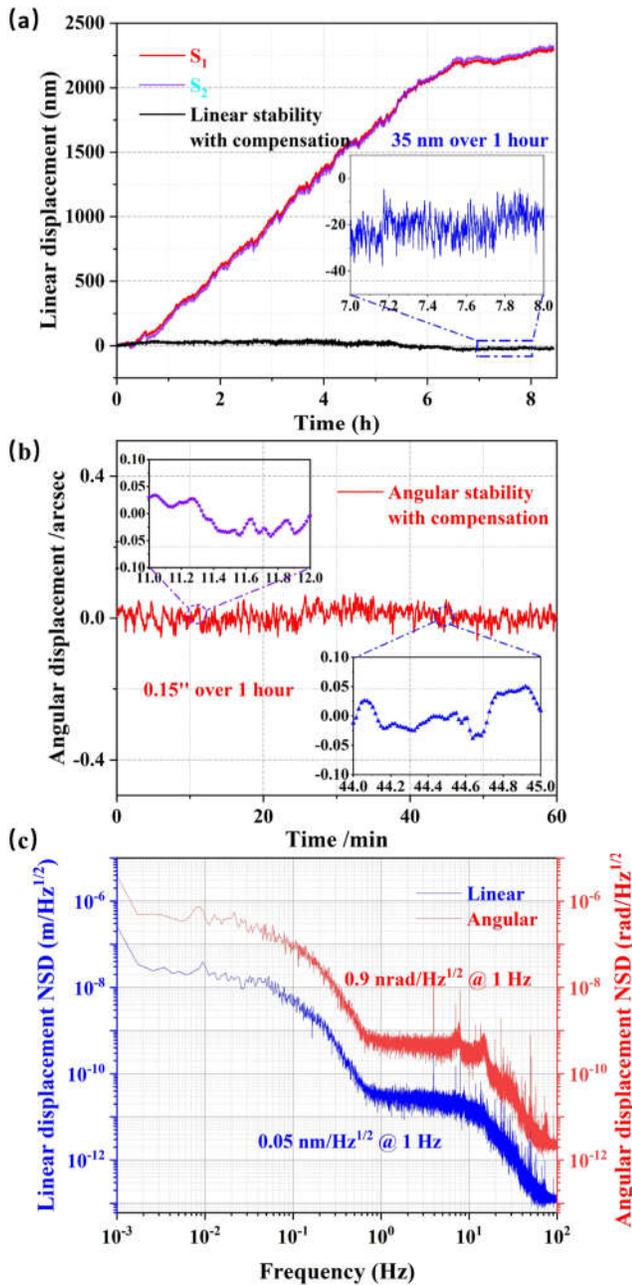

Fig. 5. Stability tests. (a) Linear displacement stability of 8 hours. (b) Angular displacement stability of 1 hour. (c) Stability noise spectrum density, linear & angular.

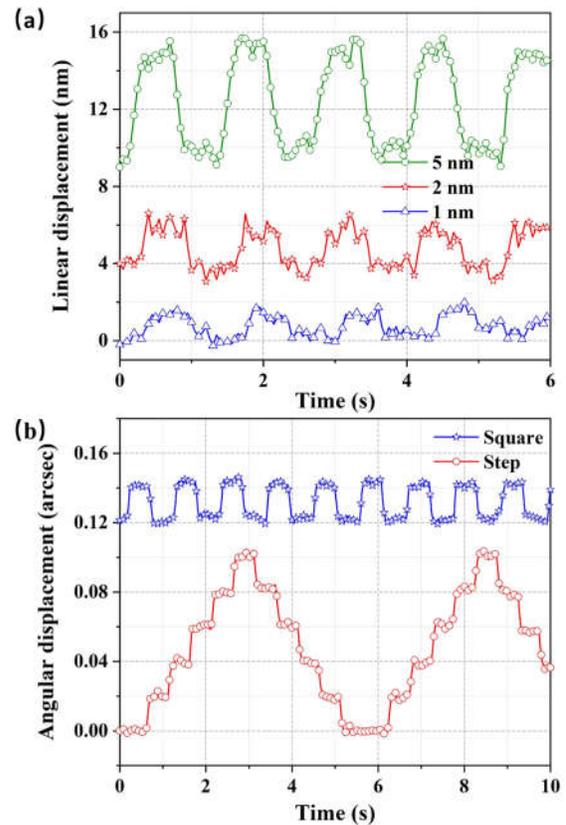

Fig. 6. Resolution test. (a) Linear displacement with 1 nm. (b) Angular displacement with 0.02''.

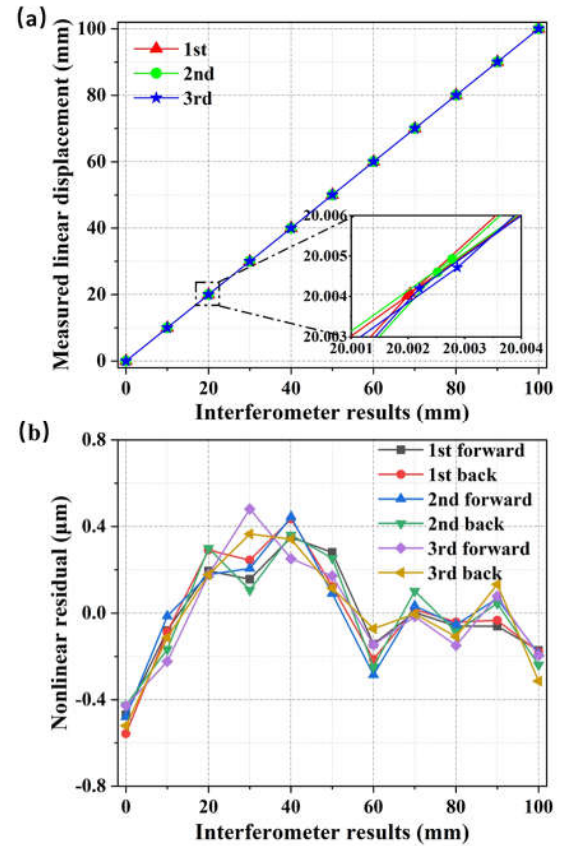

Fig. 7. Linear displacement measurement (a) Results compared with the interferometer. (b) Residual results of linear fitting.

*2. Resolution*

To determine the resolution of the prototype, a high-precision six-freedom stage (PI, Inc. P-562.6CD, 1 nm displacement resolution and 0.02" rotation resolution) is employed. The corresponding results are shown in Fig. 6. 1 nm linear displacement and 0.02" angular displacement can be distinguished using the proposed dual-beam sensor system, though the influence of disturbance drift.

*3. Linearity and Repeatability*

Linearity is one of the most important elements for the incremental displacement/rotation sensors. For linear displacement measurement, a heterodyne interferometer (Leice, Inc. LH3000, ±0.4 μm/m accuracy) is employed as the reference. The comparison experimental setup is shown in Fig. 4(a). The customized target is placed on the linear stage (PI, Inc. M511, ±0.2 μm repeatability), and the cooperative mirrors for the interferometer are also installed on the same moving stage. The measured results through the prototype are plotted in Fig. 7 compared with the interferometer results. The residual results indicate that the prototype has a nonlinearity of 0.558 μm within the 100 mm displacement range.

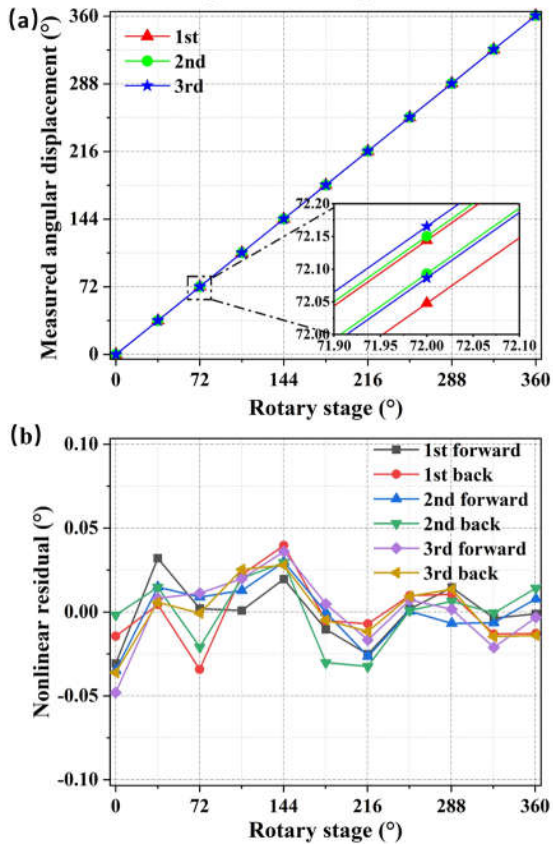

Fig. 8. Angular displacement measurement (a) Results compared with the rotary stage. (b) Residual results of linear fitting.

The angular performance is tested by a high-precision rotary stage (Beijing Optical Century Instrument, RS211, 10" precision over 360° range). A 36° step-by-step rotation forward and back in the 360° range has been launched out, and the test target is an aluminum cylinder of 10 mm diameter. The measured results are shown in Fig. 8. The angular nonlinearity is 0.048° in the 360° range under three times experiments.

Repeatability is also another important indicator for the linear/angular displacement sensors, which is defined as the standard deviation of the measured position error between the starting position and the final position. For the built prototype, it can be also calculated from the residual results. The linear repeatability is 0.28 μm and the angular repeatability is 0.023°.

The characteristics of the prototype system are summarized in Table I. Comparison of the proposed sensor system with the existing sensing schemes is presented in Table II.

TABLE I CHARACTERISTICS OF THE PROTOTYPE SENSOR SYSTEM

|         | Parameter     | Value                      |
|---------|---------------|----------------------------|
| Linear  | Range         | 0-100mm                    |
|         | Stability     | 35 nm over 1 hour          |
|         | Resolution    | 1 nm                       |
|         | Linearity     | $5.58 \times 10^{-6}$ / 100 mm |
|         | Repeatability | 0.28 μm / 100mm            |
| Angular | Range         | 0-360°                     |
|         | Stability     | 0.15" over 1 hour          |
|         | Resolution    | 0.02"                      |
|         | Linearity     | $1.34 \times 10^{-4}$ / 360°   |
|         | Repeatability | 0.023° / 360°              |

TABLE. II COMPARISON OF THE CURRENT METHODS FOR LINEAR AND ANGULAR DISPLACEMENT MEASUREMENT*

| Sensing principle | Displacement range that can be measured | | Stability | Resolution | Linearity | Mirrors or Contact to the target is required | Easy to manufacture |
|---|---|---|---|---|---|---|---|
| | Linear | Angular | | | | | |
| Capacitive [14] | No | 360° | 2" @ 15s | 0.54" | $2.78 \times 10^{-6}$ @ 360° | Yes | Moderate |
| Magnetic [16] | 300mm | No | - | - | $1.3 \times 10^{-3}$ @300mm | Yes | Moderate |
| Photodetector [25] | 500μm | 2° | - | 6nm 0.01° | $3 \times 10^{-4}$ @ 100mm $5 \times 10^{-2}$ @ 2° | No | Complex |
| Fiber interferometer | 400mm | No | 2.5nm/1h | 0.4nm | $1.5 \times 10^{-7}$ @400mm | Yes | Complex |
| Keysight 5530 Interferometer [41] | 80m | ±10° | 60nm/1h 0.40"/1h | 0.25nm 0.005" | $2 \times 10^{-8}$ @ 80m $2 \times 10^{-3}$ @ 20° | Yes | Complex |
| This paper | 100mm | 360° | 35nm/1h 0.15"/1h | 1nm 0.02" | $5.58 \times 10^{-6}$ @ 100mm $1.34 \times 10^{-4}$ @ 360° | No | Moderate |

*The parameters are concluded from the experimental results of the referred papers, which are not equal to the performance limit of listed sensors.

IV. DISCUSSION AND EXTENDED APPLICATIONS

A. Error analysis

According to the test results, the linearity is $5.58 \times 10^{-6}$ within the 100 mm range and $1.34 \times 10^{-4}$ within the 360° range, respectively. The systematic errors mainly result from the misalignment and the deviation of parameters to the real value, while the random error is caused by the parameter variation of the sensor system during the testing process. To explore the theoretical performance, uncertainty analysis for the linear and angular displacement measurement are brought out based on the sensing principle as below:

$$\frac{u(\Delta L)}{\Delta L} = \sqrt{\left(\frac{u(\lambda)}{\lambda}\right)^2 + \frac{u^2(\Delta\phi_1) + u^2(\Delta\phi_2)}{(\Delta\phi_1 - \Delta\phi_2)^2}} \quad (9)$$

$$\frac{u(\Delta\theta)}{\Delta\theta} = \sqrt{\left(\frac{u(\lambda)}{\lambda}\right)^2 + \frac{u^2(\Delta\phi_1) + u^2(\Delta\phi_2)}{(\Delta\phi_1 - \Delta\phi_2)^2} + \left(\frac{u(D)}{D}\right)^2} \quad (10)$$

where $u$ represents the uncertainty.

For linear sensing, the relative uncertainty of the laser wavelength is $1\times10^{-7}$, and the phase relative stability is $2\times10^{-6}$ within the 100 mm range. Thus, taking the confidence factor as 2, the relative measurement uncertainty of linear displacement is calculated to be $4\times10^{-6}$/100 mm. When the sensor system is used to measure the angular displacement, the distance $D$ has an uncertainty of 5 μm, which is dependent on the beam size and short-term drift. The angular relative uncertainty is calculated to be $1.48\times10^{-4}$/360° with a confidence factor of 2. The estimated theoretical uncertainty is in good agreement with the experimental results. Based on the error analysis, the future improvement of the linear/angular displacement measurement should locate in the alignment with the target, the calibration and stabilization of sensor parameters.

*B. Extended applications-PZT calibration & Rotation monitor*

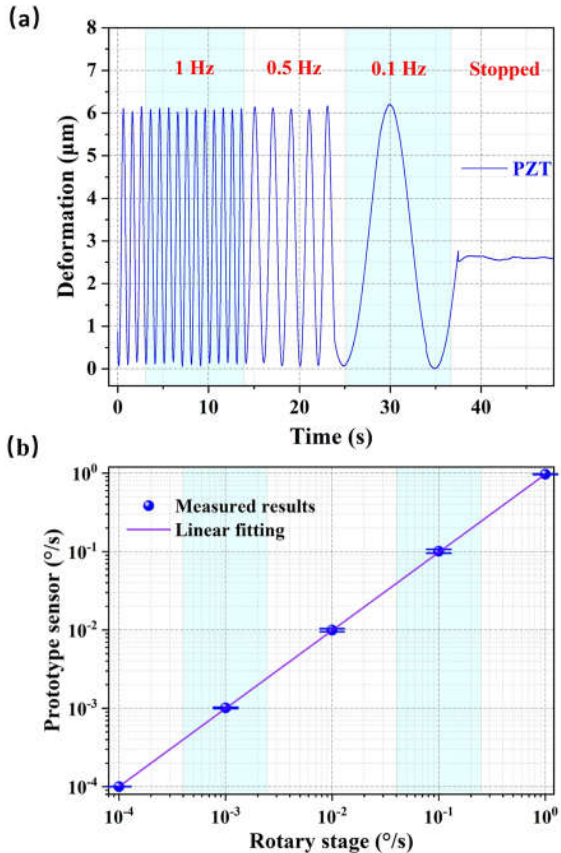

Fig. 5. Applications of the proposed sensor system. (a) Linear displacement measurement of a commercial PZT actuator. (b) Angular velocity measurement of a rotating stage with error bar.

Apart from the fundamental performance tests of the prototype system, applications with PZT testing and rotation monitoring are also accomplished, as shown in Fig. 5. A PZT actuator (CoreMorrow, Inc. XMT 150) is used as a measured target and the deformation is controlled by a signal generator (Keysight, 33600A). The micrometer-level deformation has been effectively recovered using the prototype. Note that the sensing beams are incident onto the target directly, without mirrors or contact to the target. Angular velocity can also be traced through the derivation of rotation versus time. Given that the data are discrete, the average time for the velocity calculation is 1 s and the results are presented in Fig. 5(b). The two applications demonstrated that the proposed sensor system can trace the linear and angular displacement in a noncontact way without mirrors.

Conclusion

In this paper, a novel optical approach for linear and angular displacement measurement has been developed. Owing to the high sensitivity of frequency-modulated feedback interferometry, the sensor system shows potential in many applications with its noncontact sensing feature. The performance tests demonstrate that wide-range linear/angular displacement sensing with nano-level resolution of 1 nm and 0.02" has been achieved. The stability, linearity, repeatability and applications of the linear and angular displacement are also demonstrated in the paper. Compared with other linear/angular displacement sensors, as shown in Table II, the advantages of the proposed sensor system could be concluded as follows.

1) High resolution and precision: Due to the high sensitivity of laser feedback interferometry and precise phase demodulation, high sensing resolution and precision are obtained compared with normal optical sensors.
2) Noncontact measurement: The sensor system does not require contact measurement or any installation of cooperative mirrors.
3) Wide sensing range: The proposed prototype has the ability to trace large displacement (~m level [42]) and full-circle rotation for non-cooperative targets.

These features provide an excellent solution to the noncontact applications of linear/angular displacement measurements. The tested nonlinearity of the prototype sensor system is $5.58\times10^{-6}$/100mm for linear displacements and $1.34\times10^{-4}$/360° for angular displacements. However, it is only a prototype in the laboratory. In future research, there should be more research that focuses on instrumentalization and engineering issues. The optimized system could promise a wide prospect in industrial applications, including robotics, factory automation, automotive domain, and others.